\documentclass[aps,prl,superscriptaddress,showpacs,twocolumn]{revtex4}
\usepackage{graphicx}
\usepackage{amsmath}
\begin{document}

\title{Transport Measurement of Andreev Bound States in a Kondo-Correlated Quantum Dot}

\author{Bum-Kyu Kim}
\affiliation{Korea Research Institute of Standards and Science, Daejeon 305-340, Republic of Korea}
\affiliation{Chonbuk National University, Jeonju 561-756, Republic of Korea}
\author{Ye-Hwan Ahn}
\affiliation{Korea Research Institute of Standards and Science, Daejeon 305-340, Republic of Korea}
\affiliation{Department of Physics, Korea University, 15-ka Anam-dong, Seoul 136-701, Republic of Korea}
\author{Ju-Jin Kim}
\affiliation{Chonbuk National University, Jeonju 561-756, Republic of Korea}
\author{Mahn-Soo Choi}
\affiliation{Department of Physics, Korea University, 15-ka Anam-dong, Seoul 136-701, Republic of Korea}
\author{Myung-Ho Bae}
\affiliation{Korea Research Institute of Standards and Science, Daejeon 305-340, Republic of Korea}
\author{Kicheon Kang}
\affiliation{Department of Physics, Chonnam National University, Gwangju 500-757,
 Republic of Korea}
\author{Jong Soo Lim}
\affiliation{Institut de F\'{i}sica Interdisciplinar i de Sistemes Complexos IFISC (CSIC-UIB)}
\author{Rosa L\'{o}pez}
\affiliation{Institut de F\'{i}sica Interdisciplinar i de Sistemes Complexos IFISC (CSIC-UIB)}
\affiliation{Departament de F\'{i}sica, Universitat de les Illes Balears}
\author{Nam Kim}
\affiliation{Korea Research Institute of Standards and Science, Daejeon 305-340, Republic of Korea}
\date{\today}
\begin{abstract}
We report non-equilibrium transport measurements of gate-tunable Andreev bound states in
a carbon nanotube quantum dot coupled to two superconducting leads. In
particular, we observe clear features of two types of Kondo ridges, which
can be understood in terms of the interplay between the Kondo effect and
superconductivity. In the first type (type I), the coupling is
strong and the Kondo effect is dominant. Levels of the Andreev bound states
display anti-crossing in the middle of the ridge. On the other hand, crossing of the
two Andreev bound states is shown in the second type (type II) 
together with the 0-$\pi$ transition of the Josephson junction.
Our scenario is well understood in terms of only a single dimensionless 
parameter, $k_BT_K^{min}$/$\Delta$, where $T_K^{min}$ and $\Delta$ are the 
minimum Kondo temperature of a ridge and the superconducting order parameter, 
respectively. 
Our observation is consistent with measurements of the critical
current, and is supported by numerical renormalization group calculations.
\end{abstract}
\maketitle

 In an Andreev reflection process at the interface between a normal metal (or any other non-
 superconducting region) and a superconductor (S), an electron in the normal region is converted to a Cooper
pair necessitating a reflection of a hole. Multiple Andreev reflections (MAR) play a central role in finite-bias
transport through a non-superconducting region sandwiched by two superconducting leads \cite{jung11,takayanagi95,du08,buitelaar03,xiang06}.
Contrary to MAR peaks at finite bias, Andreev bound states (ABS) are
formed as a result of coherent superposition of all possible Andreev reflection
processes (to the infinite number). 
So far ABS have been observed either in equilibrium across 
two superconducting electrodes \cite{pillet10}, or in a system with
only one superconducting lead~\cite{dirks11,deacon10}.
The interplay of ABS with the Kondo effect in S-quantum dot (QD)-S is an interesting issue, 
particularly in the
context of the 0-$\pi$ transition~\cite{glazman89,spivak91,choi04,siano04,sellier05,vandam06}.
The ``$\pi$ state", exhibiting the reversed sign of the Josephson current of S-QD-S,
originates from an unpaired electron spin in a strongly interacting QD~\cite{glazman89,spivak91}. The usual ``0 state" is
recovered due to the Kondo screening in the strong coupling limit. 
Theoretically, it has been shown that this 0-$\pi$ transition is followed by the level crossing
of two ABS ~\cite{lim08}.
Interestingly, the nature of the ground state is switched between the ``0-state" and the ``$\pi$-state" at the ABS crossing point.
However, simultaneous observation of ABS level crossing and the 0-$\pi$
transition has never been achieved experimentally.

 In this Letter, we report a clear signature of a gate-tunable ABS in non-equilibrium transport through a
carbon nanotube QD asymmetrically coupled to two superconducting leads. 
In particular, we show that a non-equilibrium transport measurement
probes the ABS (which is regarded as an equilibrium property) together
with the  0-$\pi$ transition in Josephson critical current ($I_c$) measurement.
Further, we find that the interplay between
ABS and the Kondo correlation plays a major role in transport. 
This leads to the two
different prototypes of the Kondo ridges depending on the ratio
$k_BT_K^{min}$/$\Delta$.
$T_K^{min}$ and $\Delta$  represent
the minimum Kondo temperature of the ridge and the superconducting order parameter, respectively. ``type-I" Kondo ridge with a stronger coupling
($k_BT_K^{min}/\Delta \gtrsim 0.8$) displays an anti-crossing of ABS,
where the Josephson junction is always in the ``0 state". On the other hand,
``type-II" Kondo ridge with a weaker coupling ($k_BT_K^{min}/\Delta \lesssim 0.8$)
shows a crossing of ABS, which is directly related to the $0$-$\pi$  transition of the Kondo-correlated Josephson junction. 
Our transport data clearly show two types of the Kondo ridge in a single sample. 
This scenario is confirmed by our measurement of gate-dependent
Josephson critical current $I_c$ along the ridge. In addition, our experimental 
result is supported by numerical renormalization group (NRG) calculations.

{\em Experimental setup.} - Carbon nanotubes (CNTs) are grown by
the conventional chemical vapor
deposition method on SiO$_2$/Si (500 nm/ 500 $\mu$m) wafer~\cite{kong98}.
A CNT is located by atomic force
 microscopy. Contact electrodes (10 nm/80 nm Ti/Al bilayer) are subsequently
 realized by electron-beam lithography and successive electron beam evaporation
 processes \cite{kim10}. Ti layer is used as an adhesion layer between the
 superconducting Al layer and CNT. The superconducting transition temperature $T_c$ of
 the bilayer is about 1.1 K. All measurements are performed at a base temperature
 below 100 mK in a dilution refrigerator. Two terminal DC measurements are done
both in the current and in the voltage bias modes by using DC voltage/current source (Yokogawa GS200), as
 shown in Fig. 1(a) \cite{com}.
 Al superconducting electrodes are switched to normal state by applying an external magnetic field of $\sim$ 1 kG at the base temperature.

\begin{figure}
\includegraphics[width=3in]{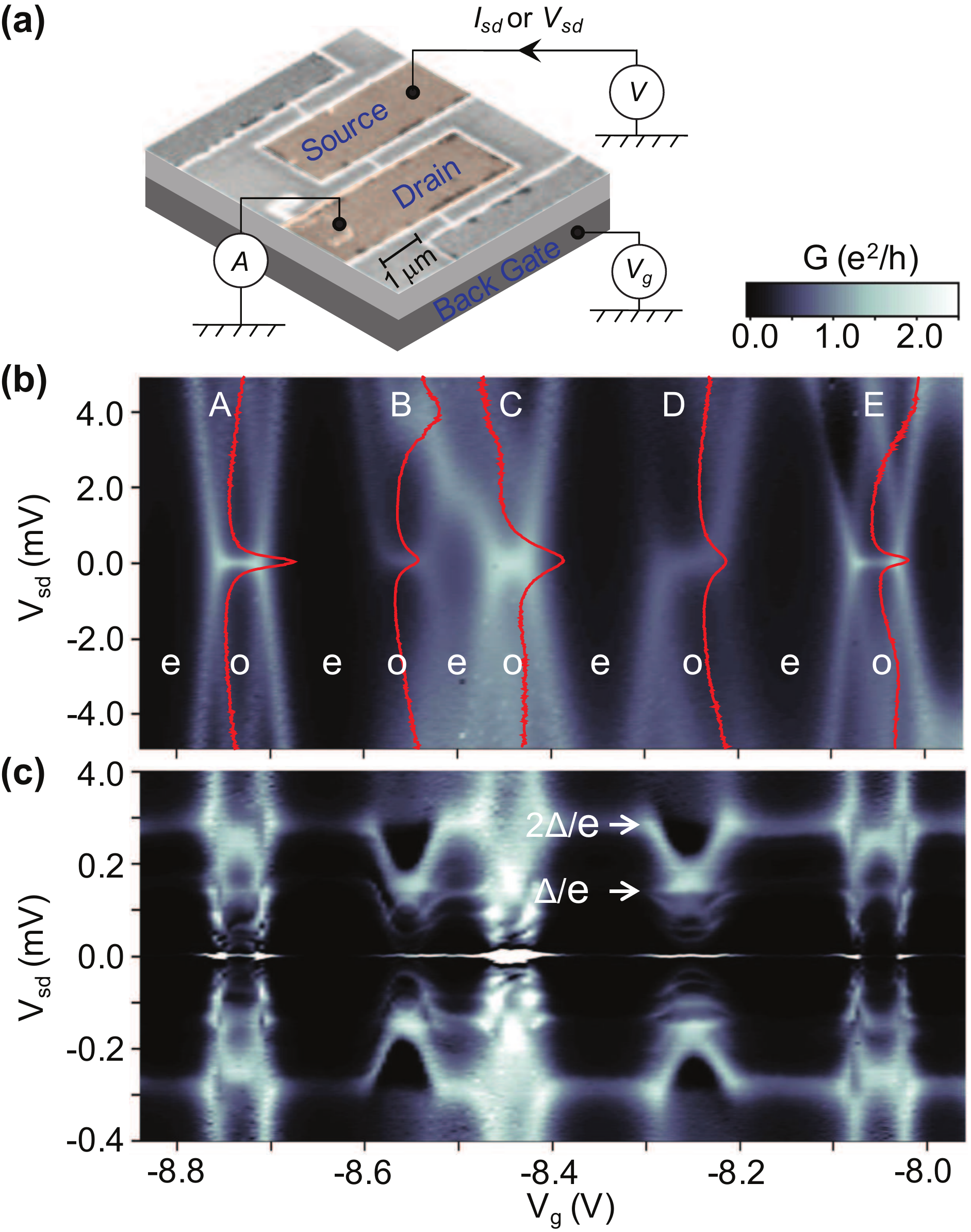}
\caption{(a) Schematic of the measurement configuration. Highly doped Si wafer and SiO$_2$ layer are used as a  back gate and
an insulating barrier, respectively. Channel length of the CNT is designed to be 300 nm. Diameter of the CNT $\sim$1.5 nm is measured by AFM. Differential conductance $dI/dV_{sd}$ plot as a function of $V_{sd}$ and $V_g$ is displayed (b) for  normal (with magnetic field B $\sim$ 1 kG), and (c) for superconducting states at zero field, respectively. The letters `e' and `o' denote the even and odd number states, respectively. Bias voltages corresponding to
$\Delta/e$ and $2\Delta/e$ (with $\Delta \sim 140$ $\mu$eV) are
indicated by arrows in (c).
Red curves in (b) correspond to $dI/dV_{sd}$ vs. $V_{sd}$ in the middle of the Kondo
ridges A - E.}
\label{fig1}
\end{figure}

The normal state differential conductance $dI/dV_{sd}$ is plotted in Fig. 1(b), providing a typical even-odd
behavior with pronounced Kondo peaks at the zero bias of odd-number ridges.
From Fig.~1(b), we extract a charging energy $U$ $\sim$ 3 meV and a level separation ${\delta}E$ $\sim$ 3 meV. The Kondo temperature $T_K$, estimated from the half width half maximum of the Kondo peaks~\cite{buitelaar02},
depends on the gate voltage. The minimum Kondo temperature $T_K^{min}$ in the
middle of each Kondo ridge is estimated to be 0.94, 1.6, 2.6, 2.4 and 0.86 K for
the Kondo ridges A to E, respectively. The tunnel coupling
$\Gamma (=\Gamma_R+\Gamma_L)$  (with  tunnel
coupling to the right(left) electrode $\Gamma_R$($\Gamma_L$)) is estimated by
fitting $T_K$ data to the formula $k_BT_K=(U\Gamma/2)^{1/2}$exp$[-\pi[|4\varepsilon^2-U^2|]/8U\Gamma]$, where $\varepsilon$ is the energy level in QD tuned by gate voltages~\cite{tsvelick78,eichler09}.
The estimated values of $\Gamma$ range from 0.35 $\sim$ 0.96 meV for ridges A-E. The asymmetry ratio $\gamma (=\Gamma_R/\Gamma_L)$
is obtained from the relation
${G}_{max}=(2e^2/h)\cdot4\Gamma_R\Gamma_L/(\Gamma_R+\Gamma_L)^2$ \cite{glazman-ng88}, where ${G}_{max}$ is the zero-bias linear conductance obtained in the middle
of each Kondo ridge (not shown here)~\cite{asymmetry}. The estimated $\gamma$ 
is 2.7, 16, 2.3, 12, and 3.1 for ridges A - E, respectively. This asymmetry 
plays an important role in ABS-assisted transport, as we discuss below.

  {\em Main features of the finite-bias transport.} -
Differential conductance with the superconducting leads is displayed in Fig. 1(c).
While the even number state shows the gate-independent (elastic) quasiparticle
cotunneling ($eV_{sd}=\pm2\Delta$) \cite{rasmussen09} together with a weak single
Andreev reflection ($eV_{sd}=\pm\Delta$) peaks, the odd-number state displays a
rich subgap structure ($|eV_{sd}|< 2\Delta$), which originates from the interplay between the Kondo effect and superconductivity \cite{buitelaar02}.
The most prominent feature is the two different prototypes of the Kondo ridges.
This can be seen more clearly in Fig. 2(a), a magnified view of ridges D and E
in Fig. 1(c). Fig. 2(a) displays a strong subgap transport.
We focus on the main peaks ($V_{sd}=V_{ABS}$) which vary from $|eV_{ABS}|=\Delta$ to $|eV_{ABS}|=2\Delta$
as a function of the gate voltage (blue dashed lines).
   Such strong peaks cannot be understood in terms of the perturbative MAR peaks.
Notably, this gate-dependent peak in the Kondo ridge
evolves into the elastic quasi-particle cotunneling peak at $|eV_{sd}|=2\Delta$ in the even valley.

\begin{figure}
\includegraphics[width=3.0in]{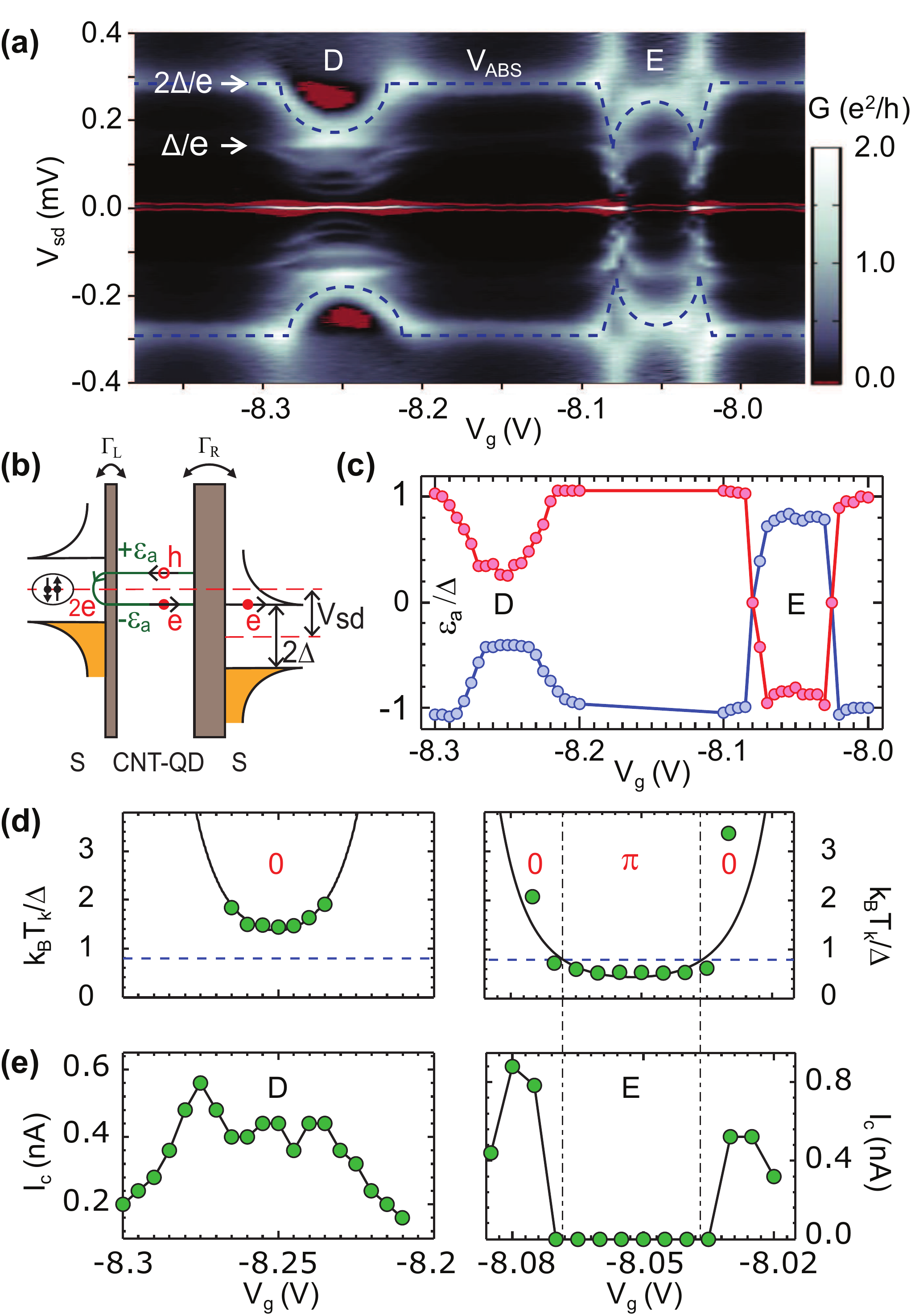}
\caption{(a) Differential conductance plot for the two types of the
Kondo ridge (magnified view of Fig. 1(c) for ridge D and E). Red color represents
the region of the negative differential conductance. Blue dashed lines
($V_{ABS}$) are associated with the  ABS-assisted transport (see the text).
(b) Schematic diagram of the major transport process forming the Andreev bound
states at finite bias. This mechanism is expected to be pronounced in highly
asymmetric barriers, where the Andreev bound states are pinned to the lead with
stronger coupling.
(c) The ABS level position $\varepsilon_a$ (in unit of $\Delta$) obtained from Eq.~(1)  is plotted as
a function of $V_g$.
(d) $k_BT_K/\Delta$ vs. $V_g$ for ridges D (type-I) and E (type-II), respectively.
Filled circles and solid lines correspond to the experimental data and the
theoretical fit with the formula
$k_BT_K=(U\Gamma/2)^{1/2}$exp$[-\pi[|4\varepsilon^2-U^2|]/8U\Gamma]$, and
$U/\Delta = 21.4$. The dashed line at $k_BT_K/\Delta =0.788 $ is the
theoretically expected `0-$\pi$ transition' line \cite{sellier05}.
(e) $I_c$ vs. $V_g$ for ridges D and E, respectively. Solid lines are for
an eye-guide. $I_c$-$V_g$ curves for ridge E show sharp drop around the
0-$\pi$ transition point denoted by the two vertical dot-dashed lines.
}
\label{fig2}
\end{figure}

We attribute these main peaks to ABS-assisted transport, which is illustrated in Fig. 2(b). In highly asymmetric junctions, ABS are formed mainly between QD and a lead with lower barrier (left lead in Fig. 2(b)). The other lead with higher barrier would play a role of probing ABS. In this picture, the gate-dependent bright peaks at $V_{sd}$ $=$ $V_{ABS}$ result from the alignment of ABS and the gap edge of the ``probe" lead. Together with the fact that ABS is formed always in pair with the electron-hole symmetry~\cite{lim08}, this gives the following relation between ABS energy ($\varepsilon_a$) and the peak position ($V_{ABS}$):
\begin{equation}
 \varepsilon_a = \pm(|eV_{ABS}|-\Delta).
\end{equation}

The gate-dependent ABS obtained from this relation is plotted in Fig. 2(c).
ABS displays two distinct features depending on the type of the Kondo
ridges. Type-I ridge (ridge D) with a stronger coupling displays an anti-crossing
in the middle of the ridge where the Kondo temperature has its minimum value.
On the other hand, type-II ridge (ridge E) shows two clear crossing points of
ABS, which is related to the 0-$\pi$ transition. The two different types are
determined by the ratio $k_BT_K^{min}$/$\Delta$. In type-I ridge, this ratio is
always larger than the critical value ($k_BT_K^{min}$/$\Delta$ $\simeq$ 0.8)
(left panel of Fig. 2(d)). In contrast, the Kondo temperature in type-II ridge has two transition points corresponding to 0-$\pi$ transition
in the right panel of Fig. 2(d)).

  {\em Critical currents } - Our scenario is further supported by the behavior of
the Josephson critical current, $I_c$, (Fig. 2(e))~\cite{icmeasure}.
High Josephson current flows due to the Kondo-assisted transmission in the strong coupling ridge (type I), and the system is always in ``0 state" (left panel of Fig. 2(e)). In type-II ridge, 0-$\pi$ transition leads to a dramatic change in the behavior of $I_c$ (right panel of Fig. 2(e)). As the Kondo effect is suppressed in the middle of the ridge, ``$\pi$ state" appears, and it leads to a strong suppression of $I_c$. All these features are consistent with the behavior of the gate-dependent ABS of Fig. 2(c). This behavior of the supercurrent has also been reported in Ref.~\cite{eichler09}, without delving it into ABS in finite-bias transport.

  {\em Numerical results} - Our main observation in Fig. 2 is nicely supported
 by NRG calculation. Fig. 3(a) shows the calculated ABS level 
$\varepsilon_a/\Delta$ as a function of the QD energy level $\varepsilon_d$. 
It clearly displays two different prototypes, i.e. anti-crossing (left) and 
crossing (right) of ABS for parameter values of $\Gamma/\Delta$ $=$ 4.9 
and 3.0, respectively (these values correspond to the experimentally extracted 
values for ridge D and E, respectively). Theoretical values of $I_c$ are 
determined by the amplitude of the current-phase relation at zero bias voltage.
Fig. 3(b) shows the QD-level dependence of $I_c$ for the two types of the 
ridge. As in the experimental result of Fig. 2(e), large Josephson current is 
assisted by the Kondo effect for type-I ridge. Suppression of the supercurrent 
is clearly shown for type-II ridge at ``$\pi$ state" (right panel of Fig.~3(b)).
This sudden drop of the supercurrent takes place exactly at the 0-$\pi$ 
transition point. All the features in the calculations match perfectly with 
the experimental observation in Fig.~2~\cite{nrg}.

 {\em Universality of ABS-assisted transport} -  We note that the gate-independent peaks for even number state at $eV_{sd}$ $=$ $\pm2\Delta$ can also be understood within our framework of ABS-assisted transport. Of course, these peaks can be identified as the elastic cotunneling of quasiparticles~\cite{rasmussen09}. On the other hand, ABS are pinned to $\varepsilon_a$ $=$ $\pm\Delta$ in the off-resonance limit of the even valley. In this case,
ABS-assisted transport also provides peaks at $eV_{sd}$ $=$ $\pm2\Delta$, according to Eq.~(1). That is, ABS peaks evolve into the quasi-particle cotunneling peaks in the off-resonance limit. Therefore, the main features of transport can be understood in a very universal way, ranging from the Coulomb blockade (off-resonance), the intermediate coupling ($\pi$ state), to the strong Kondo limit (0 state).

\begin{figure}
\includegraphics[width=3in]{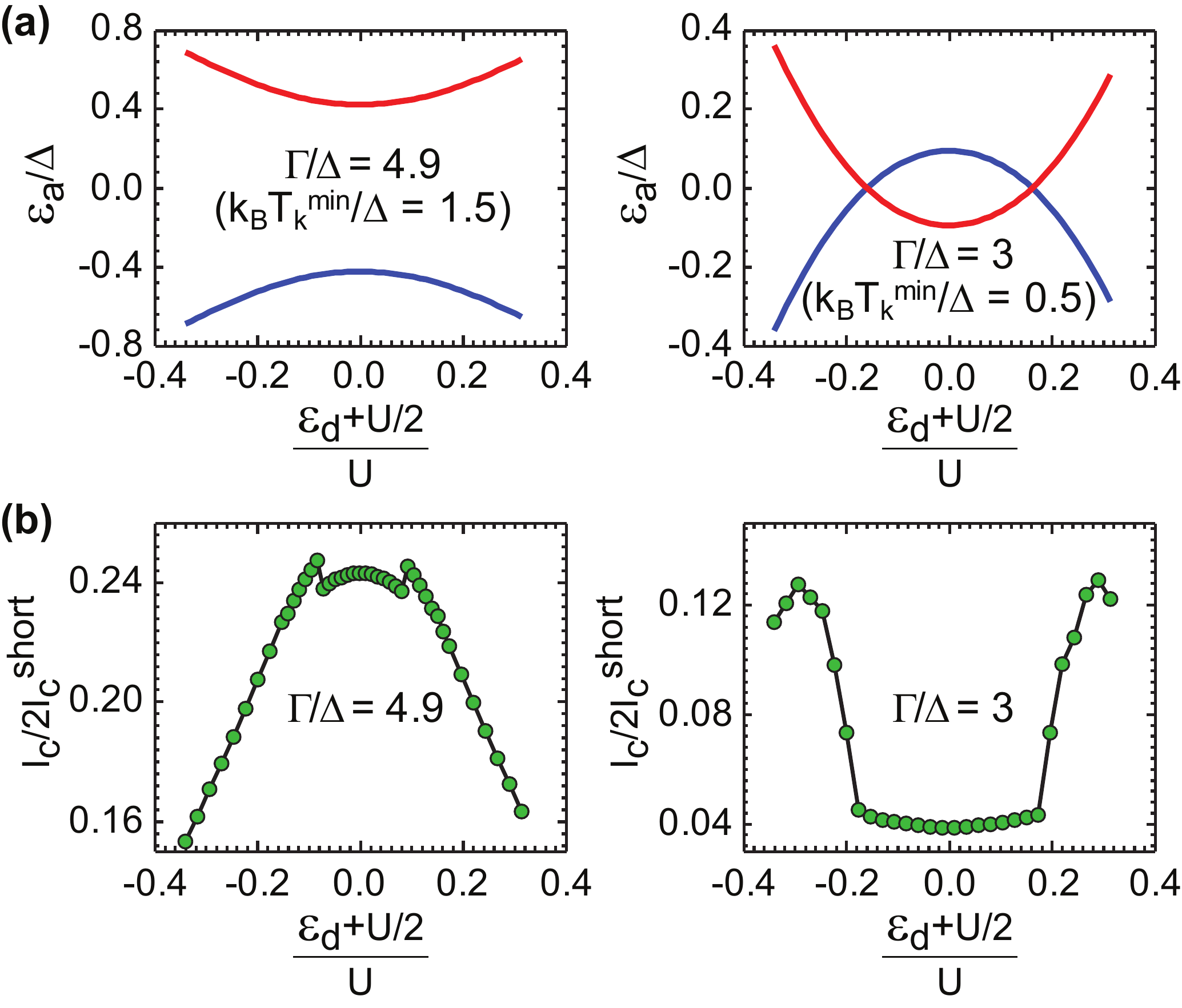}
\caption{NRG results for the two types of the Kondo ridge: (a) ABS level position $\varepsilon_a$ (in unit of $\Delta$) and (b) the Josephson critical current $I_c$ as a function of the QD energy level $\varepsilon_d$. Both in (a) and (b), coupling constants $\Gamma/\Delta$ $=$ 4.9 (left panels) and $\Gamma/\Delta$ $=$ 3.0 (right panels) are used, which correspond to the experimental values of the ridges D (type-I) and E (type-II), respectively.
Experimentally extracted value of $U/\Delta$$=21.4$ is used in both cases. $k_BT_K^{min}$/$\Delta$ are 1.5 and 0.5 for ridge D and E, respectively.}
\label{fig3}
\end{figure}

  {\em Negative dynamic conductance and fine structures} - Another evidence of
ABS-assisted transport is shown in the negative dynamic conductance (NDC) region
in the voltage bias mode (blue line in Fig. 4). It can be more clearly seen
in $dI/dV_{sd}$ (upper-left inset of Fig. 4 and the red-colored region in
Fig.~2(a)). Hysteretic switching current above $I_c$ is found in a current bias at
$I=I_a$ (forward) or $I=I_{ar}$ (backward) depending on the direction of the
current sweep. NDC and hysteretic switching current can be understood as a general
feature when Cooper pair tunnels through a resonant state in a
QD~\cite{nevirkovets07,andersen11,yeyati97}. For highly asymmetric barriers
(as illustrated in Fig. 2(b)), resonant ABS formed with the lower barrier
lead can be probed by the ``probe" lead. However, as tunneling barriers become
more symmetric, ABS levels are not well defined at finite bias and NDC is expected to disappear.
Actually, NDC and the hysteretic switching current at $I=I_a$ and $I=I_{ar}$
are observed only for ridge B and D whose asymmetry ratio is very large
($\gamma=16$ and $\gamma=12$ for ridges B and D,
respectively). When one of the two tunneling barriers is significantly higher
than the other, our measurement configuration is equivalent to scanning tunneling
microscope (STM) measurement with a superconducting tip. Thus, the dynamic
conductance density plots for ridge B and D reflect the density
of states in the CNT-QD as shown in Refs.~\cite{pillet10,dirks11,deacon10}.

Finally, we briefly discuss the fine structures in the differential conductance at lower voltage, $|eV_{sd}|$ $<$ $\Delta$ in Fig.~1(c) and Fig.~2(a). Although a quantitative analysis for this is beyond our scope, we notice that the smaller peaks at $|eV_{sd}|<\Delta$ greatly resemble the gate dependence of the main peaks at $V_{sd}$ $=$ $V_{ABS}$ for both types of the Kondo ridge. Therefore, we speculate that these small peaks originate from the combination of ABS (mainly formed with the lower barrier contact) and single or multiple Andreev reflections (with the higher barrier contact).

\begin{figure}
\includegraphics[width=2.8in]{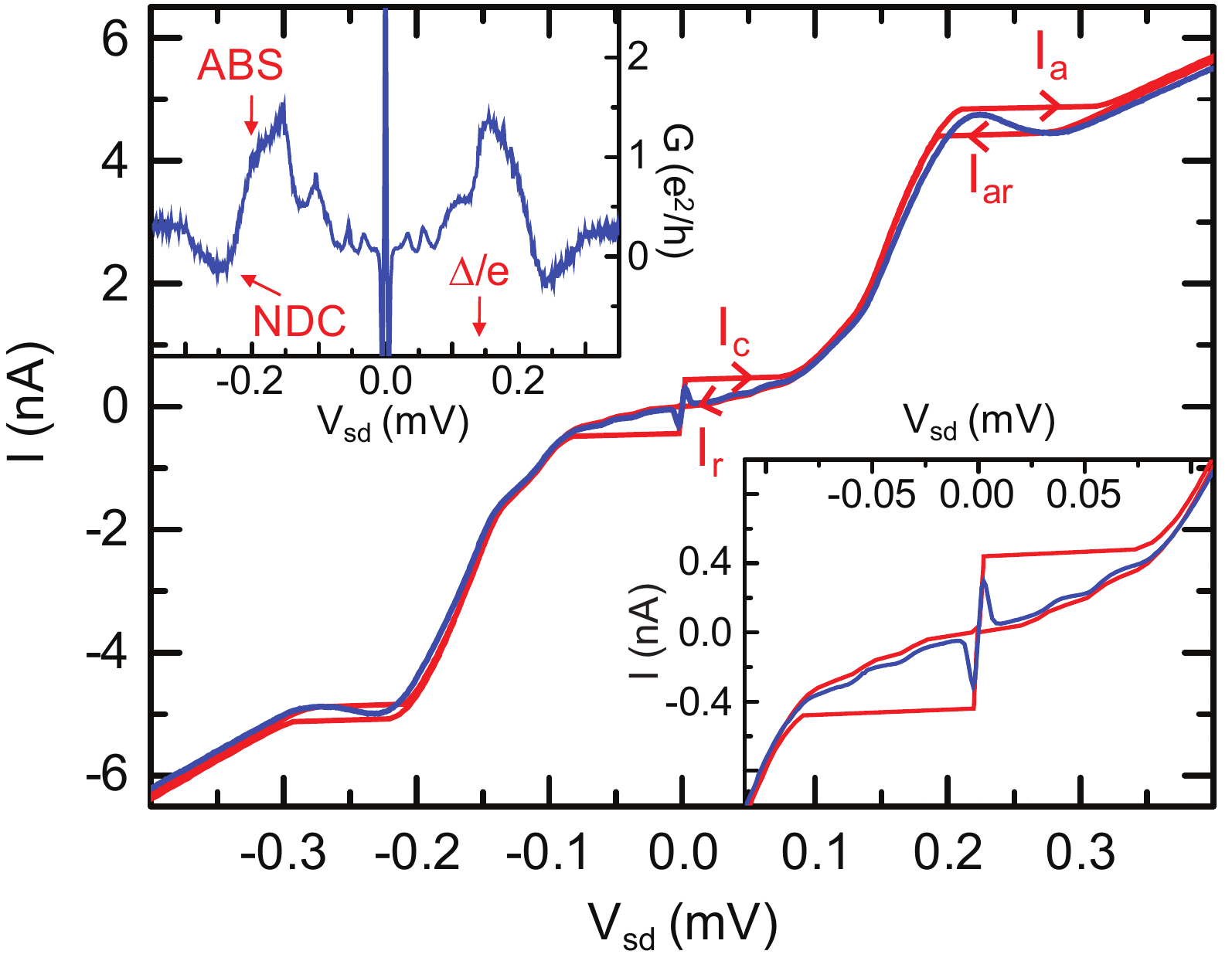}
\caption{$I-V_{sd}$ curves measured in a current bias (red line), and voltage bias  (blue line) modes, respectively, in the middle of the Kondo ridge D ($V_g$=-8.255 V). Two switching currents, $I_c$ and $I_a$, appear at zero voltage and at $V\approx0.2$ mV, respectively.  The slope of the supercurrent branch is mainly due to the resistance of low pass filters ($\sim$ 4 k$\Omega$). Upper-left inset: differential conductance $G = dI/dV_{sd}$ vs. $V_{sd}$  in voltage bias mode.
Lower-right inset: magnified view of the $I$-$V$ curve around zero bias.}
\label{fig4}
\end{figure}

 {\em Conclusion} - In conclusion, gate tunable ABS are reported in $I$-$V$ measurement configuration in an Al-CNT-Al Josephson junction. The observed differential conductance shows the two distinct types of the Kondo ridges associated with ABS. ABS displays crossing (anti-crossing) behavior, which is the main characteristics of the 0-$\pi$ transition (0 junction) tuned by a gate voltage applied to the QD. This feature is also consistent with a measurement of the gate-dependent critical current, and is confirmed by a numerical renormalization group calculation.

\begin{acknowledgments}
We acknowledge valuable discussions with Yong-Joo Doh and thank
Hyun-Ho Noh
and Woon Song for assistance in measurement.
This work was supported by by the National Research Foundation of Korea (NRF)
(Grant No.2009-0084606, No.2012R1A1A2003957, 2010-0025880, 2011-0012494 and
No.2011-0015895), and by the Korea Research Institute of Standard and Science.
R.~L and J.S.~L were supported by MINECO Grants Nos. FIS2008-00781, FIS2011-23526
and CSD2007-00042 (CPAN).

\end{acknowledgments}


\begin{thebibliography}{99}
%
\bibitem{jung11} M. Jung, H. Noh, Y.-J. Doh, W. Song, Y. Chong, M.-S. Choi, Y. Yoo, K. Seo, N. Kim, B.-C. Woo, B. Kim, and J. Kim, ACS Nano~{\bf 5}, 2271 (2011).

\bibitem{takayanagi95} H. Takayanagi and T. Akazaki, Phys.~Rev.~B~{\bf 52}, R8633 (1995).

\bibitem{du08} X.~Du, I.~Skachko, and E.~Y.~Andrei, Phys.~Rev.~B~{\bf 77}, 184507 (2008).


\bibitem{buitelaar03} M. R. Buitelaar, W. Belzig, T. Nussbaumer, B. Babic, C. Bruder, and C. Schonenberger, Phys.~Rev.~Lett.~{\bf 91}, 057005 (2003).

\bibitem{xiang06} J. Xiang, A. Vidan, M. Tinkham, R. M. Westervelt, and C. M. Lieber, Nature Nanotech.~{\bf 1}, 208 (2006).

\bibitem{pillet10} J-D. Pillet, C. H. L. Quay, P. Morfin, C. Bena, A. Levy Yeyati and P. Joyez, Nat. Phys.~{\bf 6}, 965 (2010).


\bibitem{dirks11} T. Dirks, T. L. Hughes, S. Lal, B. Uchoa, Y.-F. Chen, C. Chialvo, P. M. Goldbart and N. Mason, Nat.~Phys.~{\bf 7}, 386 (2011).

\bibitem{deacon10} R. S. Deacon, Y. Tanaka, A. Oiwa, R. Sakano, K. Yoshida, K. Shibata, K. Hirakawa, and S. Tarucha, \prl{\bf 104}, 076805 (2010).

\bibitem{glazman89} L.~I.~Glazman and K.~A.~Matveev, Pis'ma~Zh. Tekh. Fiz. {\bf 49}, 570 (1988)
 [JETP~Lett.~{\bf 49}, 659 (1989)].
\bibitem{spivak91} B.~I.~Spivak and S.~A.~Kivelson, Phys.~Rev.~B~{\bf 43}, 3740 (1991).

\bibitem{choi04} M.-S. Choi, M. Lee, K. Kang, and W. Belzig, \prb~{\bf 70}, 020502 (2004).

\bibitem{siano04} F. Siano and R. Egger, \prl~{\bf 93}, 047002 (2004).

\bibitem{sellier05} G. Sellier, T. Kopp, J. Kroha, and Y. S. Barash, \prb~{\bf 72}, 174502 (2005)

\bibitem{vandam06} J. A. van Dam, Y. V. Nazarov, E. P. A. M. Bakkers, S. De Franceschi,and L. P. Kouwenhoven, Nature {\bf 442}, 667 (2006); J.-P. Cleuziou, W. Wernsdorfer, V. Bouchiat, T. Ondarcuhu and M. Monthioux, Nature Nanotech. {\bf 1}, 53 (2006); H. I. J{\o}rgensen, T. Novotny, K. Grove-Rasmussen, K. Flensberg,and P. E. Lindelof, Nano Lett. {\bf 7}, 2441 (2007).

\bibitem{lim08} J.~S.~Lim and M.-S.~Choi, J.~Phys.~Cond.~Mat~{\bf 20}, 415225 (2008); J. Bauer, A. Oguri and A. C. Hewson, J. Phys.: Cond. Mat. {\bf 19}, 486211 (2007).

\bibitem{kong98} J. Kong, H. T. Soh, A. M. Casse, C. F. Quate, and H. Dai, Nature {\bf 395}, 878 (1998).

\bibitem{kim10} B.-K. Kim, J.-J. Kim, M. Seo, Y. Chung, B.-C. Woo, J. Kim, W. Song, and N. Kim, Appl. Phys. Lett. {\bf 97}, 262110 (2010).

\bibitem{com} In order to suppress high frequency noise from heating sample, low-pass RC filters with a cut-off frequency of $\sim$ 10 kHz are mounted on the sample chip. The line resistance including filters in two-terminal configuration is measured to be $\sim$ 4 k$\Omega$. Low pass filters are essential to observe the switching current in small magnitude.

\bibitem{buitelaar02} M. R. Buitelaar, T. Nussbaumer, and C. Schonenberger, \prl~{\bf 89}, 256801 (2002).

\bibitem{tsvelick78} A. M. Tsvelick and P. B. Wiegmann, Adv. Phys. {\bf 32}, 453 (1983); F. D. M. Haldane, Phys. Rev. Lett. {\bf 40}, 416 (1978).

\bibitem{eichler09} A. Eichler, R. Deblock, M. Weiss, C. Karrasch, V. Meden, C. Schonenberger, and H. Bouchiat, Phys. Rev. B {\bf 79}, 161407 (2009).

\bibitem{glazman-ng88} G.~L.~I. Glazman and M.E.~Raikh, Pis’ma Zh. \'{E}ksp. Teor. Fiz.~{\bf 47}, 378 (1988) [JETP Lett.~{\bf 47}, 452 (1988)];
  T.~K.~Ng and P.~A.~Lee, Phys. Rev. Lett.~{\bf 61}, 1768 (1988).


\bibitem{asymmetry} This is possible because every Kondo ridge is almost in the unitary limit ($T\ll T_K$) in our base temperature.

\bibitem{rasmussen09} K. Grove-Rasmussen, H. I. Jorgensen, B. M. Andersen,J. Paaske, T. S. Jespersen, J. Nyg\aa rd, K. Flensberg, and P. E. Lindelof, Phys. Rev. B {\bf 79}, 134518 (2009).
\bibitem{icmeasure} $I_c$ is measured from the switching current in the $I-V$
curve as shown in Fig.4.
%
\bibitem{nrg} Symmetric coupling is considered in our calculation. Note that
the asymmetry $\gamma$ does not modify the position of ABS, and gives only a
gate-independent reduction of $I_c$.

%
\bibitem{nevirkovets07} I. P. Nevirkovets, S. E. Shafranjuk, O. Chernyashevskyy, and J. B. Ketterson, Phys. Rev. Lett. {\bf 98}, 127002 (2007).

\bibitem{andersen11} B. M. Andersen, K. Flensberg, V. Koerting, and J. Paaske, Phys. Rev. Lett. {\bf 107}, 256802 (2011).

\bibitem{yeyati97} A. Levy Yeyati, J. C. Cuevas, A. Lopez-Davalos, and A. Martin-Rodero, Phys. Rev. B {\bf 55}, R6137 (1997).



\end{thebibliography}
\end{document}